\documentclass[twocolumn,floatfix,prb,aps,showpacs,superscriptaddress]{revtex4-2}
\usepackage{graphicx,amsmath,amssymb,color,nicefrac,multirow, makecell, ragged2e,float,multirow,arydshln}
\usepackage[unicode=true,colorlinks=true,linkcolor=blue,citecolor=blue,urlcolor=blue]{hyperref}
\usepackage[titletoc,title]{appendix}

\newcommand{\be}{\begin{equation}}
\newcommand{\ee}{\end{equation}}

\newcommand{\ba}{\begin{eqnarray}}
\newcommand{\ea}{\end{eqnarray}}

\renewcommand{\vec}[1]{{\bm{#1}}}

\def\beq{\begin{eqnarray}}
\def\eeq{\end{eqnarray}}

\makeatletter
\newcommand*{\rom}[1]{\expandafter\@slowromancap\romannumeral #1@}
\makeatother

\newcommand{\non}{\nonumber\\}
\newcommand{\ua}{\uparrow}
\newcommand{\da}{\downarrow}
\newcommand{\Kua}{K\!\!\ua}
\newcommand{\Kda}{K\!\!\da}
\newcommand{\Kpua}{K'\!\!\ua}
\newcommand{\Kpda}{K'\!\!\da}
\newcommand{\pa}{\partial}
\usepackage{bm,physics,enumerate,booktabs,ulem}

\newcommand{\tx}[1]{{\text{#1}}}
\newcommand{\mo}{moir\'e }
\newcommand{\para}{\parallel}

\begin{document}
\title{
Quantum Anomalous, Spin, and Valley Hall Effects in Pentalayer Rhombohedral 
Graphene Moir\'e Superlattices\\
}

\author{Koji Kudo}
\affiliation{Department of Physics, Kyushu University, Fukuoka 819-0395, Japan}

\author{Ryota Nakai}
\affiliation{Department of Physics, Kyushu University, Fukuoka 819-0395, Japan}
\affiliation{RIKEN Center for Quantum Computing (RQC), Wako, Saitama, 351-0198, Japan}

\author{Kentaro Nomura}
\affiliation{Department of Physics, Kyushu University, Fukuoka 819-0395, Japan}

\begin{abstract}
Recent experiments on pentalayer rhombohedral graphene \mo superlattices have 
observed the quantum anomalous Hall effect at \mo filling factor of $\nu = 1$ 
and various fractional values. These phenomena are attributed to a flat Chern 
band induced by electron-electron interactions. In this study, we demonstrate 
that at $\nu = 2$, many-body effects can lead to the emergence of quantum spin 
Hall and quantum valley Hall states, in addition to the quantum anomalous Hall 
state, even in the absence of spin-orbit coupling or valley-dependent 
potentials. These three topological states can be selectively induced by the 
application and manipulation of a magnetic field. Furthermore, we show that at 
$\nu = 3$ and $4$, the ground state 
can be a combination of 
topologically trivial and nontrivial states,
unlike the cases of $\nu=1$ and 2.
This contrasts with the conventional quantum Hall effect in
graphene where the ground state at filling factor $\nu$ is given as the
particle-hole counterpart at $4-\nu$.
\end{abstract}

\maketitle

\section{Introduction}
The recent discovery of both integer and fractional quantum anomalous Hall 
(QAH) effects in \mo materials represents a major milestone in condensed matter
physics~\cite{Cai23,Zeng23,Park23,Xu23,ZLu24}. 
Moir\'e materials provide a fertile ground for exploring strong interaction 
effects~\cite{Andrei20, Mak22}, leading to the emergence of various 
symmetry-broken phases
such as superconductivity and Wigner 
crystals~\cite{Cao18,Aaron19,Li21,Jin21,Anderson23},
as well as topological 
phases~\cite{Dean13,Eric18,Liu19B,Kang24,Kang24B,Liu24B}. 
The QAH effect is particularly intriguing because it arises from the interplay 
between symmetry breaking and topology, resulting in a quantized Hall 
conductance even in the absence of external magnetic fields~\cite{Serlin20, Yujun20, Li21B, Benjamin24}.
Both integer and fractional QAH effects were first observed in twisted bilayer 
MoTe$_2$~\cite{Cai23, Zeng23, Park23, Xu23}, marking a significant achievement 
in the study of \mo materials.
More recently, these phenomena have been reported in pentalayer rhombohedral
graphene on hexagonal boron nitride (hBN) at a \mo filling factor of $\nu=1$ 
and at various fractional values~\cite{ZLu24}. This breakthrough has spurred 
extensive theoretical investigations~\cite{
Reddy23,Wang23B,JDong-CFL23,Goldman23,Reddy23B,Xu23B,Wang23C,Yu23,
Dong23,ZhouYang23,JDong23,Guo23,Kwan23,HArbeitman24,Liu24,Zeng24,Tan24,Soejima24,Dong24,Xie24,Sheng24,Lu24,Nuckolls24}.

The emergence of Chern bands  
in pentalayer rhombohedral graphene on hBN can be attributed to the combined effects of layer stacking, \mo structure, and many-body 
interactions~\cite{Dong23, ZhouYang23, JDong23, Guo23, Kwan23}.
 The band touching at the charge-neutral point of multilayer rhombohedral graphene ~\cite{Zhang11,Zhou21,Zhou21b,Zhou22,Barrera22,Zhang23,Han23,Han24,Liu23b,Das24,Han24B,Braz24,JXie24}
becomes increasingly flat (approximately $\sim k^{N_L}$ with wave number $k$) as the number of stacked layers $N_L$ increases~\cite{Guinea06}.
The \mo superlattice structure, resulting from the lattice mismatch between pentalayer graphene and hBN~\cite{Moon14, Jung14B}, reduces the graphene's Brillouin zone (BZ) into \mo BZ, narrowing the band width. Ultimately, 
electron-electron interactions play a crucial role in isolating a Chern band from the rest of the \mo bands.
Specifically, Hartree-Fock (HF) analysis has revealed the emergence of an isolated and nearly flat Chern band at $\nu=1$~\cite{Dong23, ZhouYang23, JDong23, Guo23, Kwan23}, which 
does not appear in the absence of electron interactions. This 
mechanism is believed to underlie the observed $\nu = 1$ and fractional QAH states~\cite{ZLu24, Dong23, ZhouYang23, JDong23, Guo23,Kwan23}.

Traditionally, quantum Hall physics~\cite{Klitzing80, Laughlin81, Thouless82, Tsui82, Laughlin83, Geim07, Novoselov05, Zhang05, Nomura06, Apalkov06, Toke06, Du09, Bolotin09, Haldane88, Neupert11, Sheng11, Regnault11, Sun11, Liu24B} relies on {\it non-interacting} topological bands, such as Landau levels, as a foundation.
However, pentalayer rhombohedral graphene on hBN 
deviates from this noninteracting framework,
indicating the potential for interaction-driven exotic phases beyond the conventional quantum Hall paradigm, even at integer fillings.
This observation prompts an investigation into pentalayer rhombohedral graphene on hBN at integer fillings other than $\nu=1$, which have not yet been explored experimentally under a sufficiently strong displacement field~\cite{ZLu24}. 

\begin{figure}[t]
\includegraphics[width=\columnwidth]{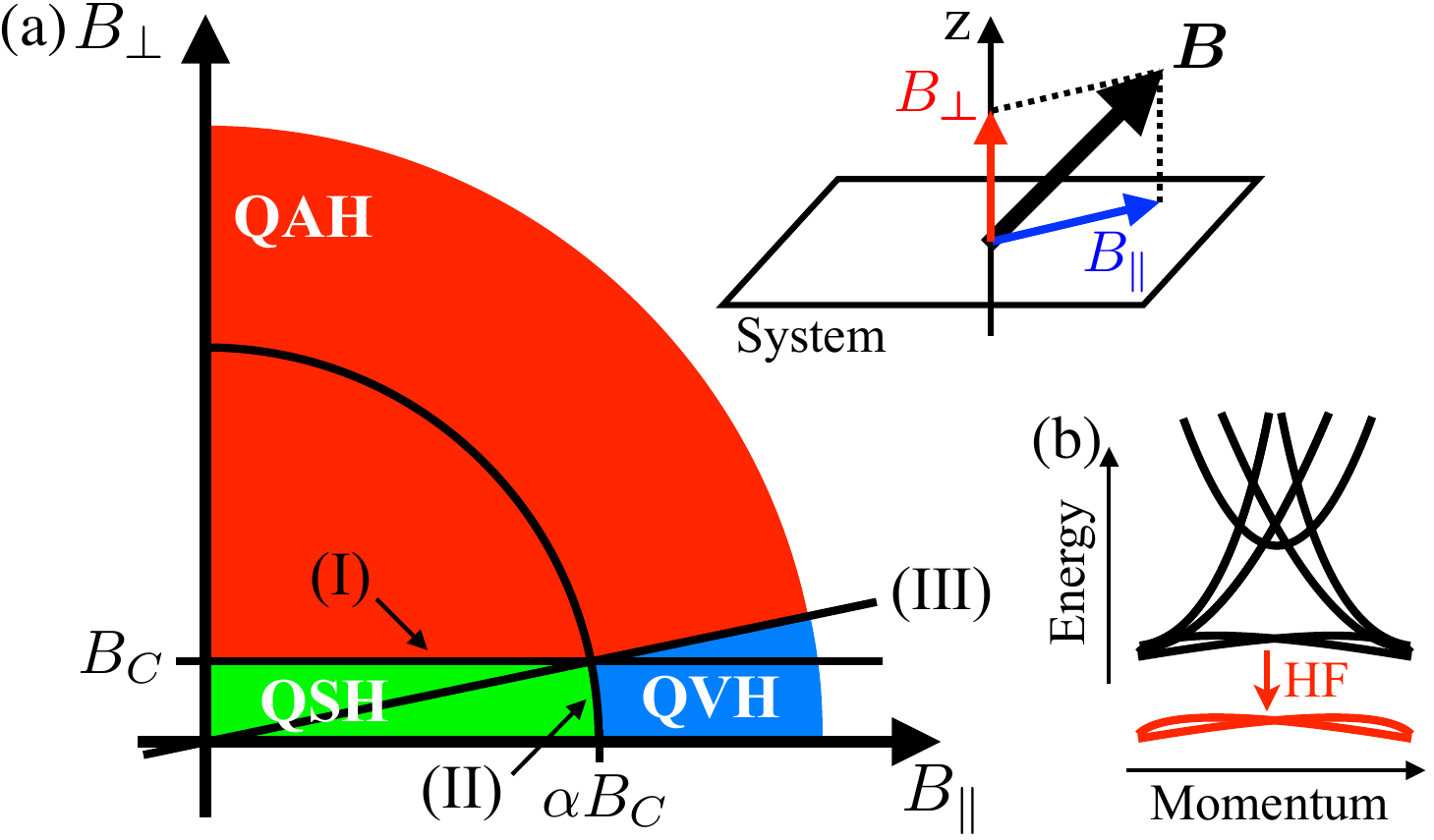}
 \caption{
 (a) Lowest energy state as a function of the in-plane and out-of-plane components $(B_\para,B_\perp)$ of the magnetic field $\vec{B}$. The boundaries are defined as follows: 
(I) QAH versus QSH transition occurs at $B_\perp=B_C$, where $B_C$ is the critical magnetic field such that $B_C \times M_\text{orb}^{z\text{(QAH)}}$, with $M_\text{orb}^z$ defined in Eq.~\eqref{eq:Morb}, equals the magnetostatic energy of the QAH state.
(II) QVH versus QSH boundary at $|\vec{B}|=\alpha B_C$ with $\alpha=4.9$ [refer to Eq.~\eqref{eq:alpha}].
(III) QVH versus QAH transition described by $\alpha B_\perp=|\vec{B}|$.
(b) Schematic of the band structure with and without the Hartree-Fock (HF) 
interaction at 
$\nu=2$. The HF interaction results in two isolated bands, as 
 demonstrated in Fig.~\ref{fig:band} below.
 The choice among the valley and spin -- $(\Kua,\Kda,\Kpua,\Kpda)$ --
 does not influence the total energy. All possible occupancies are summarized in Table~\ref{tab:nu2candidates}.
 }
 \label{fig:LE}
\end{figure}

In this study, we demonstrate that pentalayer rhombohedral graphene at $\nu = 2$ can give rise to quantum spin Hall (QSH) and quantum valley Hall (QVH) states, in addition to the QAH state, even in the absence of spin-orbit coupling or valley-dependent potentials. 
These states emerge purely from many-body effects, 
which is a kind of a topological Mott insulator
proposed in Ref.~\cite{Raghu08}.
Our self-consistent HF calculations show that the QAH, QSH, and QVH states at $\nu=2$ 
are degenerate but distinguishable by their magnetization magnitudes. This distinction allows for the selective induction of one of these states by applying and tilting magnetic fields, as illustrated in Fig.~\ref{fig:LE}(a). The emergence of these three states can be understood by analogy to the $\nu=1$ case: the emergent Chern band at $\nu=1$ spontaneously selects the valley $K$ or $K'$ and spin $\ua,\da$, 
where the band Chern number is +1 for $K$ and -1 for 
$K'$~\cite{Dong23, ZhouYang23, JDong23, Guo23, Kwan23}.  
At $\nu=2$, the ground state comprises two Chern bands selected from $(\Kua,\Kda,\Kpua,\Kpda)$ [see Fig.~\ref{fig:LE}(b)]. 
Different combinations of the four degrees of freedom result in distinct 
topological states, 
leading to QAH, QSH, and QVH effects. The possible 
combinations are detailed in Table~\ref{tab:nu2candidates}.

Notably, the empirical rule of forming the lowest energy state by occupying the
Chern bands does not hold at $\nu = 3$ and $\nu = 4$. At these filling factors,
the ground state can be a combination of 
topologically trivial and nontrivial states.
This occurs because 
the spatial charge density distributions of these states interlock; the localized density of the trivial state fits into the low-density regions of the topological states. 
This phenomenon, driven by many-body interactions, contrasts with the conventional quantum Hall effect in graphene where the ground state at filling 
factor $\nu$ is given as the particle-hole counterpart at $4-\nu$.

\section{Model}
For numerical analysis, an effective continuum model is used~\cite{Dong23, JDong23}. The starting point is the tight-binding Hamiltonian of pentalayer rhombohedral graphene, including both intralayer and interlayer hoppings as well as the interlayer potential difference [see Appendix~\ref{sec:R5G}]. Using a standard approach to reach the continuum limit, the effective Hamiltonian near the charge-neutral point is derived:
$H_{R5G}=\sum_{\vec{k}}\bm{c}^\dagger(\vec{k})h_{R5G}(\vec{k})\bm{c}(\vec{k})$
for a given valley and spin,
where 
$\bm{c}^\dagger=(c_{A_1}^\dagger,c_{B_1}^\dagger,c_{A_2}^\dagger,\ldots,c_{B_5}^\dagger)$ and $c^\dagger_{X_l}$ is a creation operator for the sublattice $X=A, B$ on the layer $l=1, 2,\ldots, 5$. The matrix $h_{R5G}(\vec{k})$ is ten-dimensional. System parameters are set to match experimental conditions for the $\nu = 1$ QAH effect~\cite{Park23-2, Dong23, JDong23, ZLu24}; see Appendix~\ref{sec:R5G}.

\begin{table}[t]
 \caption{
Possible occupancy of the four degrees of freedom at $\nu=2$, yielding ${}_4C_2=6$ states. The charge Chern number $C_c$ sums the band Chern numbers, whereas the spin (valley) Chern number $C_s$ $(C_v)$ represents the difference concerning the spin (valley). [Here, the band Chern number for the valley $K$ ($K'$) is $+1$ ($-1$).] The three types of states -- 
quantum anomalous Hall (QAH), quantum spin Hall (QSH), and quantum valley Hall 
(QVH) -- are distinguished. The magnetization results in Fig.~\ref{fig:alpha} 
are also summarized, with $\circ$ indicating finite magnetization.
Note that our QSH state also supports the QVH effect~\cite{Islam16}.
 }
 \label{tab:nu2candidates}
 \centering
  \begin{tabular*}{\columnwidth}{@{\extracolsep{\fill}}cccccccccc}
   \toprule
   & \multicolumn{4}{c}{Occupancy} & & & \multicolumn{2}{c}{Magnetization} \\
   \cmidrule{2-5}\cmidrule{8-9}
   \# &
   $\Kua$ & $\Kda$ & $\Kpua$ & $\Kpda$ & $(C_c,C_s,C_v)$ & Type 
   & Orbital & Spin\\
   \cmidrule{1-9}
   \morecmidrules
   \cmidrule{1-9}
   1 &
   $\checkmark$ & $\checkmark$ & $\times$ & $\times$ & (2,0,2) & \multirow{2}{*}{QAH} 
   & \multirow{2}{*}{$\circ$} & \\
   2 &
   $\times$ & $\times$ & $\checkmark$ & $\checkmark$ & (-2,0,2) & \\
   \cmidrule{6-9}
   3 &
   $\checkmark$ & $\times$ & $\times$ & $\checkmark$ & (0,2,2) & \multirow{2}{*}{QSH} \\
   4 &
   $\times$ & $\checkmark$ & $\checkmark$ & $\times$ & (0,-2,2)& \\
   \cmidrule{6-9}
   5 &
   $\checkmark$ & $\times$ & $\checkmark$ & $\times$ & (0,0,2)& \multirow{2}{*}{QVH}
   & & \multirow{2}{*}{$\circ$}\\
   6 &
   $\times$ & $\checkmark$ & $\times$ & $\checkmark$ & (0,0,2)& \\
   \midrule
   \bottomrule   
  \end{tabular*}
\end{table}

Stacking graphene on hBN induces a \mo superlattice structure due to lattice mismatch. In this study, the twist angle is set to $0.77^\circ$ to align with the experimental conditions described in Ref.~\onlinecite{ZLu24}. Within the effective continuum model, the impact of hBN is represented by a local potential $v(\vec{r})$ within the bottom graphene layer~\cite{Moon14,Jung14B} (see 
Appendix~\ref{sec:hBN} for more details). The second quantized form of this potential is 
$V_\text{hBN}
=\sum_{\vec{k}}\sum_{m_1m_2}\tilde{\bm{c}}^\dagger
(\vec{k}+m_1\bm{G}_1+m_2\bm{G}_2)v(m_1,m_2)\tilde{\bm{c}}(\vec{k})$,
where $\tilde{\bm{c}}^\dagger=(c_{A_1}^\dagger,c_{B_1}^\dagger)$,
$\bm{G}_1$ and $\bm{G}_2$ are the \mo reciprocal lattice vectors, and
$v(m_1,m_2)$ is the Fourier coefficient of $v(\vec{r})$. 
In the numerical calculations, the summation $\sum_{\vec{k}} \sum_{m_1 m_2}$ is confined to the first and second \mo BZs. 

The effective noninteracting Hamiltonian is obtained as $H_{R5G}+V_\text{hBN}$. To incorporate many-body effects, the HF interaction is added. The original interaction term is 
$H_\text{int}
=(1/2S)\sum_{\vec{k}\vec{k}'\vec{q}}\sum_{ZZ'}V_C(\vec{q})
c^\dagger_Z(\vec{k}+\vec{q})c^\dagger_{Z'}(\vec{k}'-\vec{q})
c_{Z'}(\vec{k}')c_Z(\vec{k})$,
where $S$ is the area of the system, and $Z$ and $Z'$ represent the spin, valley, sublattice, and layer indices. The dual gate-screened Coulomb interaction used is 
$V_C(\vec{q})=e^2/(2 \epsilon_0 \epsilon_rq)\tanh(qd_s)$
for $\vec{q}\neq \vec{0}$ and $V_C(\vec{q}=\vec{0})=0$, with $d_s=25$ nm as the gate distance and $\epsilon_r=5$ as the dielectric constant~\cite{JDong23}. 
More details on the HF interaction can be found in Appendix~\ref{sec:HF}. To reduce computational cost, the calculations are performed within the subspace of the first three noninteracting conduction bands for each valley and
spin. The system with $24 \times 24$ \mo unit cell is used unless otherwise 
specified.

\section{Band structure}
The self-consistent HF calculation assumes conservation of both spin and 
valley. The spin conservation assumption is valid due to the SU(2) 
spin-rotational symmetry of the HF Hamiltonian. 
Although valley conservation is nontrivial at this moment, we assume
it based on the HF results at $\nu=1$~\cite{Kwan23}.
\begin{figure}[t]
\includegraphics[width=\columnwidth]{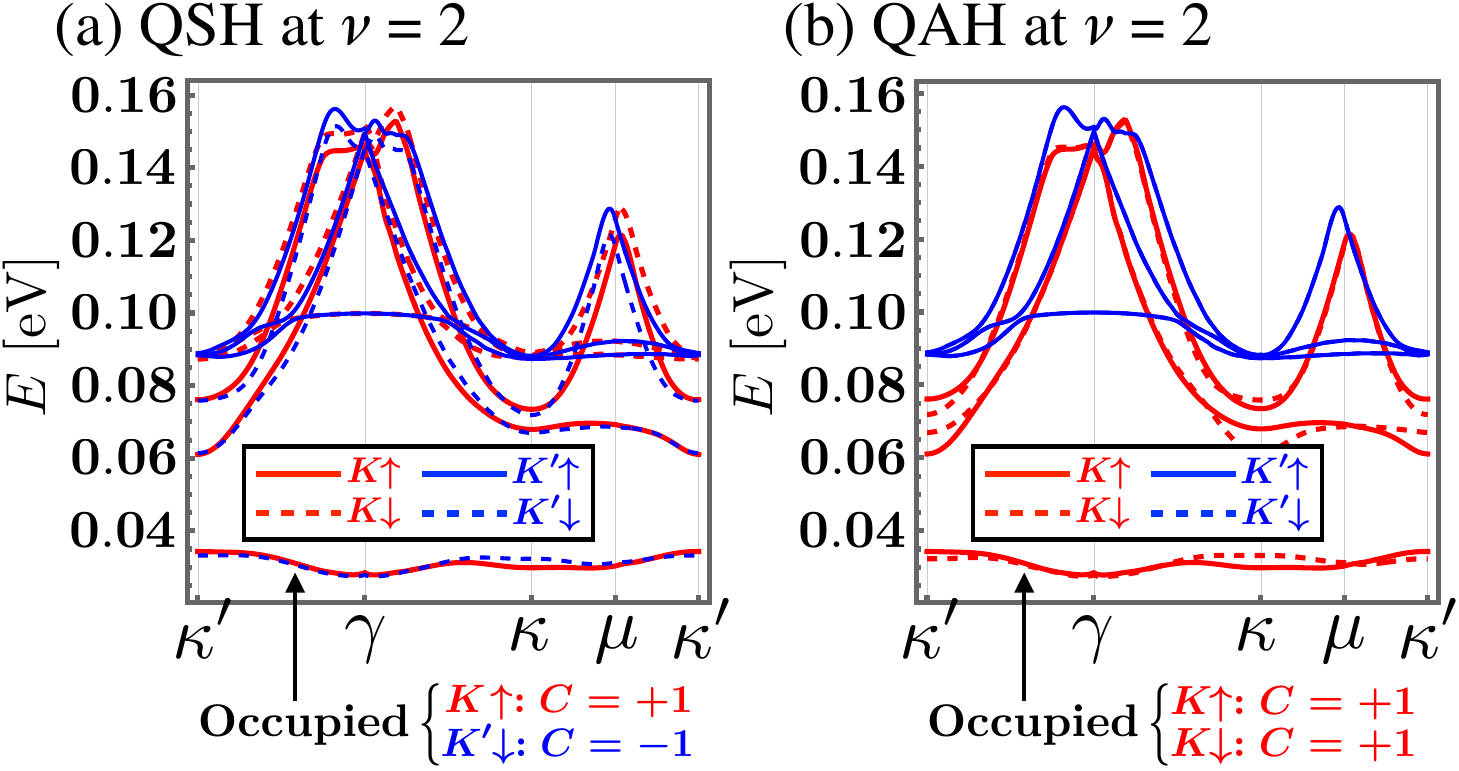}
 \caption{
Band structures at $\nu=2$ for the lowest energy Hartree-Fock solutions. 
(refer to Ref.~\cite{deg}).
Greek letters denote the high-symmetric points in the \mo BZ. Colors and line styles denote valley and spin properties. (a) and (b) correspond to the states \#3 (QSH) and \#1 (QAH) listed in Table~\ref{tab:nu2candidates}, respectively. In the calculation for (b), a small ``valley Zeeman term'' is introduced to favor valley $K$ (refer to footnote~\cite{valleyZeeman}).
The HF analysis is performed with the particle number fixed at $\nu=2$.
  }
 \label{fig:band}
\end{figure}

Figures~\ref{fig:band}(a) and (b) display the HF band structures of lowest
energy solutions, calculated by fixing the particle number at $\nu=2$. (To be
precise, the state in Fig.~\ref{fig:band}(b) has a lower energy than that in 
(a) by $~10^{-8}$eV per particle. 
Given the small difference, we consider the two states to 
be degenerate~\cite{deg}.)
In Fig.~\ref{fig:band}(a), the occupied two lowest bands, distinctly separated from other conduction bands, exhibit $(\Kua, \Kpda)$ with the band Chern numbers $C = (1, -1)$, corresponding to state \#3 (QSH) in Table~\ref{tab:nu2candidates}. In Fig.~\ref{fig:band}(b), the two lowest bands exhibit $(\Kua, \Kda)$ with band Chern numbers $C = (1, 1)$, corresponding to state \#1 (QAH) in Table~\ref{tab:nu2candidates}. Because the HF Hamiltonian maintains symmetry for spin flipping in each valley, these two results indicate that the QAH, QSH, and QVH states in Table~\ref{tab:nu2candidates} are all degenerate ground states. 
[For example, state \#5 (QVH) can be obtained by flipping the spin for $K'$ in state \#3.]
These states do not emerge in the absence of the HF interaction. Their 
emergence results from many-body effects.

\section{Lowest energy state}
We now analyze the selectivity among the QAH, QSH, and QVH states listed in Table~\ref{tab:nu2candidates} under three specific conditions: (I) In-plane magnetic field, (II) Out-of-plane magnetic field, and (III) Absence of magnetic field.

\underline{(I) In-plane magnetic field}: 
Under an in-plane magnetic field, the field favors a state with the largest in-plane magnetization, which is the QVH state due to its spin polarization. The strength of the spin magnetization for each state is quantified as follows: $M_\tx{spin}^\tx{(QVH)}=g\mu_Bn_e$ and 
$M_\tx{spin}^\tx{(QAH)}=M_\tx{spin}^\tx{(QSH)}=0$. Here, the $g$-factor is assumed to be 2, $\mu_B$ denotes the Bohr magneton, and $n_e$ represents the electron density. Note that the orbital magnetization has no in-plane component.

\begin{figure}[t]
\includegraphics[width=\columnwidth]{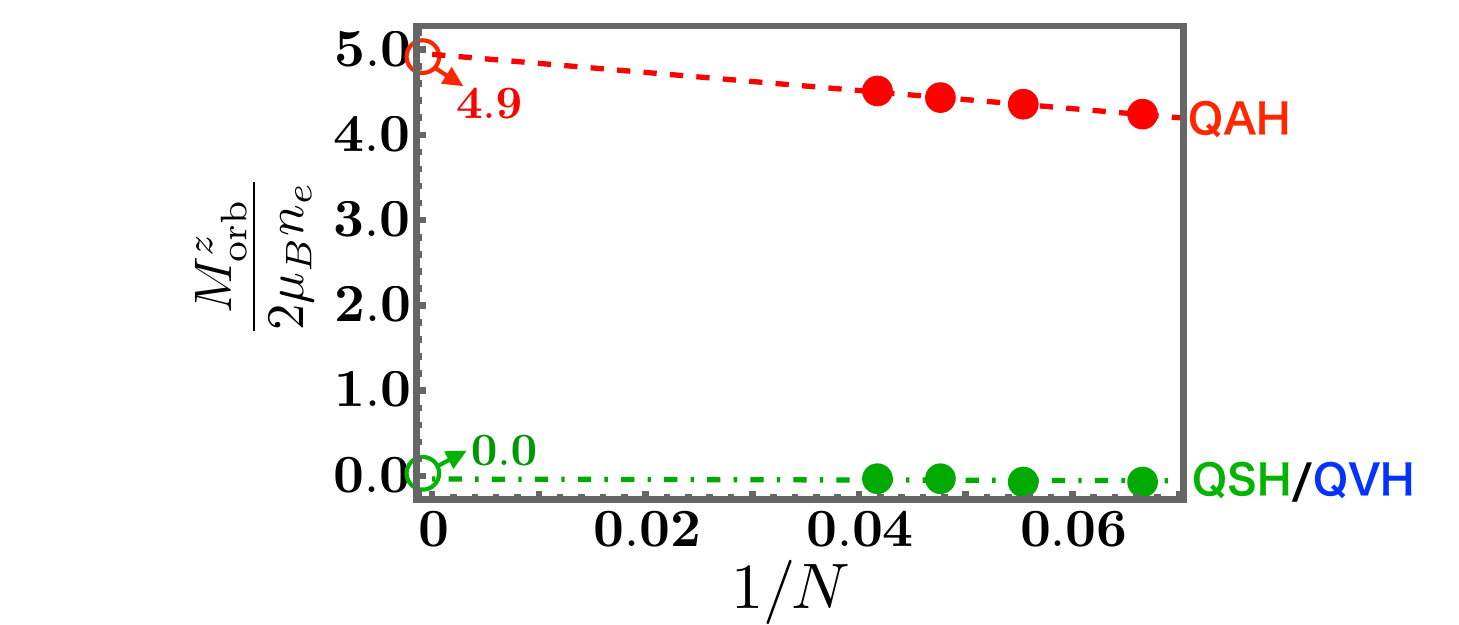}
 \caption{
Orbital magnetization $M_\text{orb}^z$, expressed in units of $2\mu_Bn_e$ where $\mu_B$ is the Bohr magneton and $n_e$ is the electron density, for $N \times N$ \mo unit cells. The states \#1 and \#3 from Table~\ref{tab:nu2candidates} serve as representatives for the QAH and QSH states, respectively. The dashed lines represent a linear approximation. 
Due to symmetry, the QSH and QVH states exhibit identical values of $M_\text{orb}^z$.
 }
 \label{fig:alpha}
\end{figure}

\underline{(II) Out-of-plane magnetic field}: 
For an out-of-plane magnetic field, both spin and orbital magnetizations are relevant. The orbital magnetization in the $z$-direction is calculated using the expression~\cite{Gat03,Xiao05,Thonhauser05,Nakai16}:
\begin{align}
 M_{\text{orb}}^z=&\frac{e}{2\hbar i}\int^\mu\frac{dk^2}{(2\pi)^2}
 \bra{\frac{\pa u}{\pa\vec{k}}}\times
 \left(H(\vec{k})+E(\vec{k})-2\mu\right)
 \ket{\frac{\pa u}{\pa\vec{k}}},
 \label{eq:Morb}
\end{align}
where $u(\vec{k})$ and $E(\vec{k})$ are an eigenstate and eigenenergy of the
Bloch Hamiltonian $H(\vec{k})$, respectively. The integral encompasses states
with energies below the chemical potential $\mu$. 
(In numerical calculations, $\mu$ is positioned at the top of the valence bands.)
In Fig.~\ref{fig:alpha}, we illustrate the orbital magnetization $M_\text{orb}^z$ for states \#1 and \#3 from Table~\ref{tab:nu2candidates}, representing the QAH and QSH states, respectively. 
Due to symmetry, the QVH state displays the same $M_\text{orb}^z$ as the QSH state. The finite-size scaling analysis presented in the figure reveals that the QSH state exhibits zero magnetization, $M_\text{orb}^{z\text{(QSH)}} + M_\text{spin}^\text{(QSH)} = 0$, while the QAH state manifests a magnetization approximately five times greater than that of the QVH state:
\begin{align}
 \frac
 {M_\text{orb}^{z\text{(QAH)}}+M_\text{spin}^\text{(QAH)}}
 {M_\text{orb}^{z\text{(QVH)}}+M_\text{spin}^\text{(QVH)}}
 \approx4.9\equiv\alpha.
 \label{eq:alpha}
\end{align}
Consequently, a QAH state is favored for the 
out-of-plane magnetic field.

\underline{(III) No magnetic field}:
In the absence of a magnetic field, the state with the smallest magnetization is favored to minimize the magnetostatic energy. Therefore, the QSH state becomes the preferred state. (At this point, it is not possible to energetically differentiate between states \#3 and \#4 in Table~\ref{tab:nu2candidates}. However, considering weak spin-orbit interactions in real systems, either of these states should be favored.)

This argument implies that the QSH state, realized in the absence of a magnetic field, transitions to the QVH or QAH states when a magnetic field is applied. Furthermore, the QVH and QAH states can be interchanged by tilting the magnetic field. 

In Fig.~\ref{fig:LE}, the state with the lowest energy is summarized as a function of both in-plane and out-of-plane components $(B_\para,B_\perp)$. The sole undetermined parameter in the figure is the critical field $B_C$, which is defined such that the product $B_C \times M_\text{orb}^{z\text{(QAH)}}$ equals the magnetostatic energy of the QAH state. Additionally, some boundaries are characterized by valley flips, representing first-order transitions. Resulting hysteresis scans of the Hall conductance is discussed in Appendix~\ref{sec:hysteresis}.

\begin{figure}[t]
\includegraphics[width=\columnwidth]{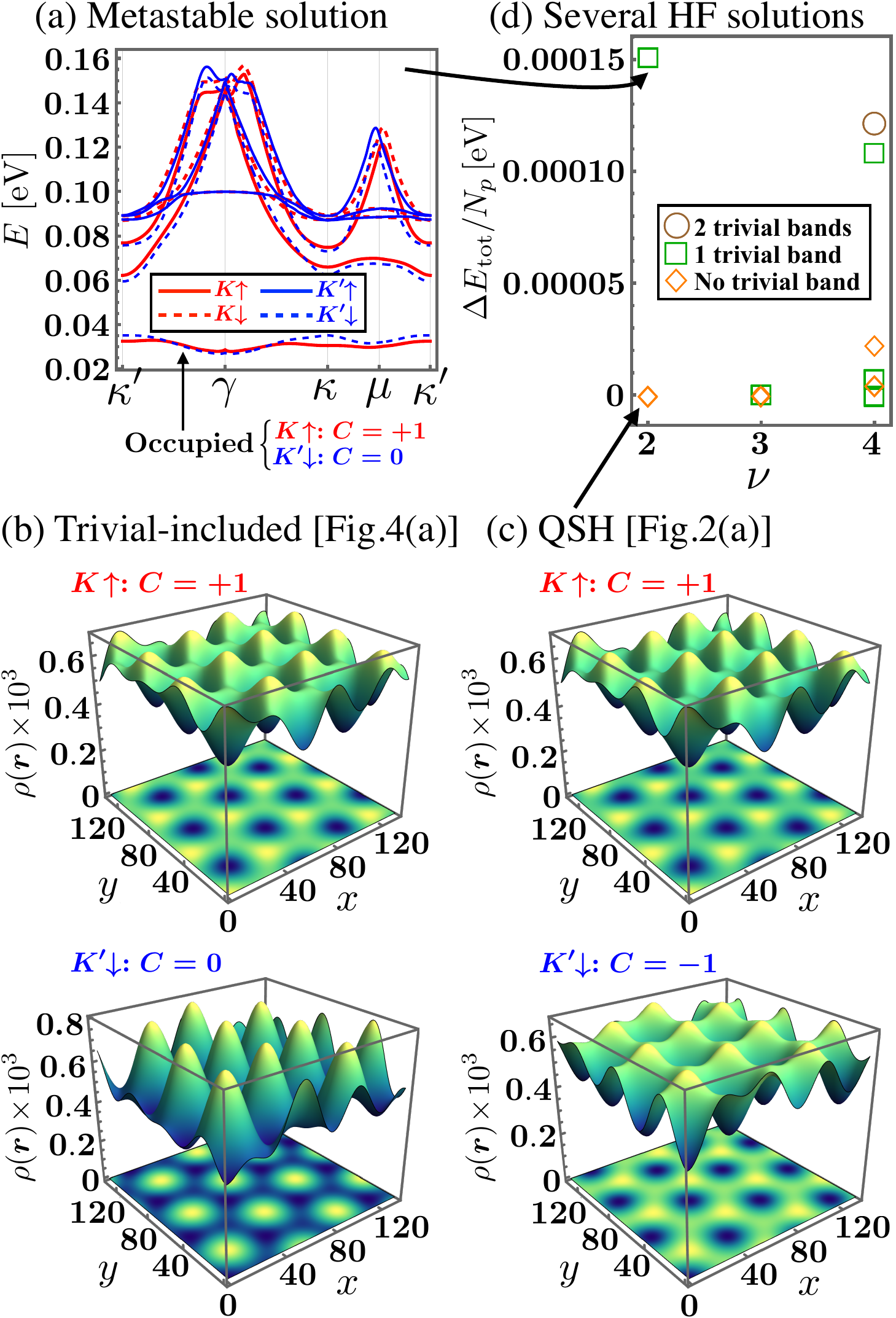}
 \caption{
(a) Hartree-Fock (HF) band structure for a metastable solution at $\nu=2$, where one of the two lowest bands is characterized by $C=0$.
(b)(c) Charge density $\rho(\vec{r})$, where the position $\vec{r}=(x,y)$ is scaled by the lattice constant of graphene $a_\text{G}$. Panels (b) and (c) correspond to the charge densities for the two lowest bands depicted in Figs.~\ref{fig:WCL}(a) and \ref{fig:band}(a), respectively. Bands with $|C|=1$ exhibit an extended structure, while that with $C=0$ displays localized densities. Light and dark colors indicate areas of high and low density, respectively.
(d) HF calculations were performed with 10-20 randomized initial states. This panel plots the difference in many-body energy per particle relative to the lowest value at each $\nu$.
(There seem to be a few plots because of the degeneracy.)
The color of the plots denotes the number of trivial bands.
 }
 \label{fig:WCL}
\end{figure}
\section{Interlocking structure in charge density}
 While the focus has been on the QAH, QSH, and QVH states, it is important to consider other competitive states.

Figure~\ref{fig:WCL}(a) depicts the HF band structure of a 
metastable solution at $\nu = 2$. The difference in many-body energy from the lowest energy state, as shown in Fig.~\ref{fig:band}(a), is about 0.00015 eV per particle. Although both band structures appear quite similar, one band in Fig.~\ref{fig:WCL}(a) has a zero band Chern number.

The two lowest bands in Fig.~\ref{fig:WCL}(a) exhibit density structures with 
an interlocking pattern.
While the densities of the topologically trivial and nontrivial bands
are qualitatively similar, the trivial band exhibits slightly more localized 
structure.
Each localized density in the trivial band fits 
into a low-density area of the Chern band.
For comparison,
Fig.~\ref{fig:WCL}(c) plots $\rho(\vec{r})$ of the QSH state from Fig.~\ref{fig:band}(a), where the two Chern bands also show 
interlocking patterns but slightly more extended structures.
Despite the seemingly more stable interlocking shape in Fig.~\ref{fig:WCL}(b), the trivial-included state is not the lowest energy state. This indicates that forming the trivial band requires more energy than the advantage gained from the interlocking structure.

\section{Lowest energy states at $\nu=3$ and 4}
Using the analogy at $\nu=2$, one might anticipate that the ground states at $\nu=3$ and $4$ would exclusively involve the Chern bands.
(Associating them with the particle-hole-like counterparts of the $\nu=1$ 
and 
$0$ ground states within the Chern band subspace, we call them PH$_{\nu=1}$ and PH$_{\nu=0}$ states, respectively). However, our findings indicate that this expectation does not always hold; the trivial-included states may achieve the lowest energy.

In Fig.~\ref{fig:WCL}(d), we present HF calculations for $\nu=2, 3, 4$ using 10-20 randomized initial states and plot the many-body energy differences relative to the lowest observed value per particle, denoted by $\Delta E_\text{tot}/N_p$. The color coding in the plots indicates the number of occupied trivial bands. The ground state at $\nu=3$ does not feature a trivial band and corresponds to the PH$_{\nu=1}$ state, yet the energy difference between this state and the trivial-included state is significantly smaller compared to that at $\nu=2$. Crucially, at $\nu=4$, the ground state includes the trivial band.

These results suggest that with increasing $\nu$, the intricate structures in the electron densities become increasingly favorable, rendering the trivial-included state at $\nu=4$ energetically more advantageous than the PH$_{\nu=0}$ state. The band structures and electron densities at $\nu=3$ and $4$ are presented in Appendix~\ref{sec:WCL34}.

\begin{table}[t]
 \caption{
Spin and orbital magnetizations, $M_\text{spin}$ and $M_\text{orb}^z$, for the low-energy states depicted in Fig.~\ref{fig:WCL}(d). 
Both magnetizations are quantified in units of $2\mu_Bn_e$.
 }
 \label{tab:nu3}
 \centering
  \begin{tabular*}{\columnwidth}{@{\extracolsep{\fill}}clccc}
   \toprule
   $\nu$ & Type & $\Delta E_\tx{tot}/N_p$ [eV] & $M_\tx{spin}$ 
		   & $|M_\tx{orb}^z|$ \\
   \midrule\midrule
   \multirow{4}{*}{3} & PH$_{\nu=1}$ & 0 & 1/3 & 1.5 \\
   \cmidrule{2-5}
   & trivial$\otimes$QAH & \multirow{3}{*}{$\lesssim10^{-6}$} & 1/3 & 2.8 \\
   & trivial$\otimes$QSH & & 1/3 & 0.3 \\
   & trivial$\otimes$QVH & & 1/3 & 0.3 \\
   \midrule
   \multirow{2}{*}{4} & trivial$\otimes$PH$_{\nu=1}$ & 0 & 0 & 0.9 \\
   \cmidrule{2-5}
   & PH$_{\nu=0}$ & $5.\times10^{-6}$ & 0 & 0.0 \\
   \midrule
   \bottomrule   
  \end{tabular*}
\end{table}

Before concluding, let us examine the phases observed at filling factors $\nu=3$ and $4$ under different conditions of magnetic fields. Table~\ref{tab:nu3} presents the spin and orbital magnetizations of the low-energy states depicted in Fig.~\ref{fig:WCL}(d). At both $\nu=3$ and $4$, the ground state of the HF Hamiltonian ($\Delta E_\tx{tot} = 0$) exhibits finite magnetization. Conversely, states with the smallest magnetization (i.e., the lowest magnetostatic energy) possess a finite $\Delta E_\tx{tot}$. Therefore, either state can be the lowest energy states in the absence of a magnetic field, depending on the ratio of $\Delta E_\tx{tot}$ to the magnetostatic energy. Upon application of an out-of-plane magnetic field of sufficient amplitude, the system transitions to the state with the largest $|M^z_\tx{orb}|$, specifically the trivial$\otimes$QAH (trivial$\otimes$PH$_{\nu=1}$) state at $\nu=3$ (4).

\section{Concluding remarks}
This study demonstrated the interplay arising as a result of many-body effects 
among three topologically distinct states -- the QAH, QSH, and QVH states -- in
pentalayer rhombohedral graphene on hBN at $\nu=2$. In the absence of a 
magnetic field, the QSH phase, characterized by zero magnetization, is 
predominant due to minimal magnetostatic energy. Application of in-plane or 
out-of-plane magnetic fields, however, shifts the preference towards the QVH 
and QAH states respectively. Therefore, manipulation of these states can be 
controlled by applying and orienting a magnetic field accordingly. 
In this work, we ignore the effects of sublattice asymmetric short-range
electron-electron and electron-phonon interactions~\cite{Kharitonov12B,Kharitonov12C,Nomura09}. It is tempting to ask
how these interactions lift the degeneracies of the QAH, QSH, and QVH states.
Additionally, we have demonstrated that the inclusion of the trivial state 
facilitates the formation of the lowest energy state at $\nu=3$ and $4$. 
This allows for the emergence of multiple topological phases.

Recent propositions suggest that composite fermions (CFs)~\cite{Jain89,Jain07} form in twisted bilayer MoTe$_2$ even in the absence of an external magnetic field~\cite{Goldman23,JDong-CFL23}. 
Generally, the CF theory establishes a mapping between the multi-component 
fractional and integer quantum Hall effects~\cite{Wu93,Scarola01b,Jain07}. If 
the CF picture holds valid in 
pentalayer rhombohedral graphene on hBN, our findings might significantly 
advance the understanding of fractional QAH physics.

\begin{acknowledgments}
 We acknowledge the computational resources offered by Research Institute for
 Information Technology, Kyushu University.
 The work is supported in part by 
 JSPS KAKENHI Grant nos. JP23K19036, 
 JP24K06926, 
 JP20H01830, 
 and JST CREST Grant
 no. JPMJCR18T2. 
\end{acknowledgments}

\appendix

\section{Pentalayer rhombohedral graphene}
\label{sec:R5G}
\begin{figure}[t]
\includegraphics[width=\columnwidth]{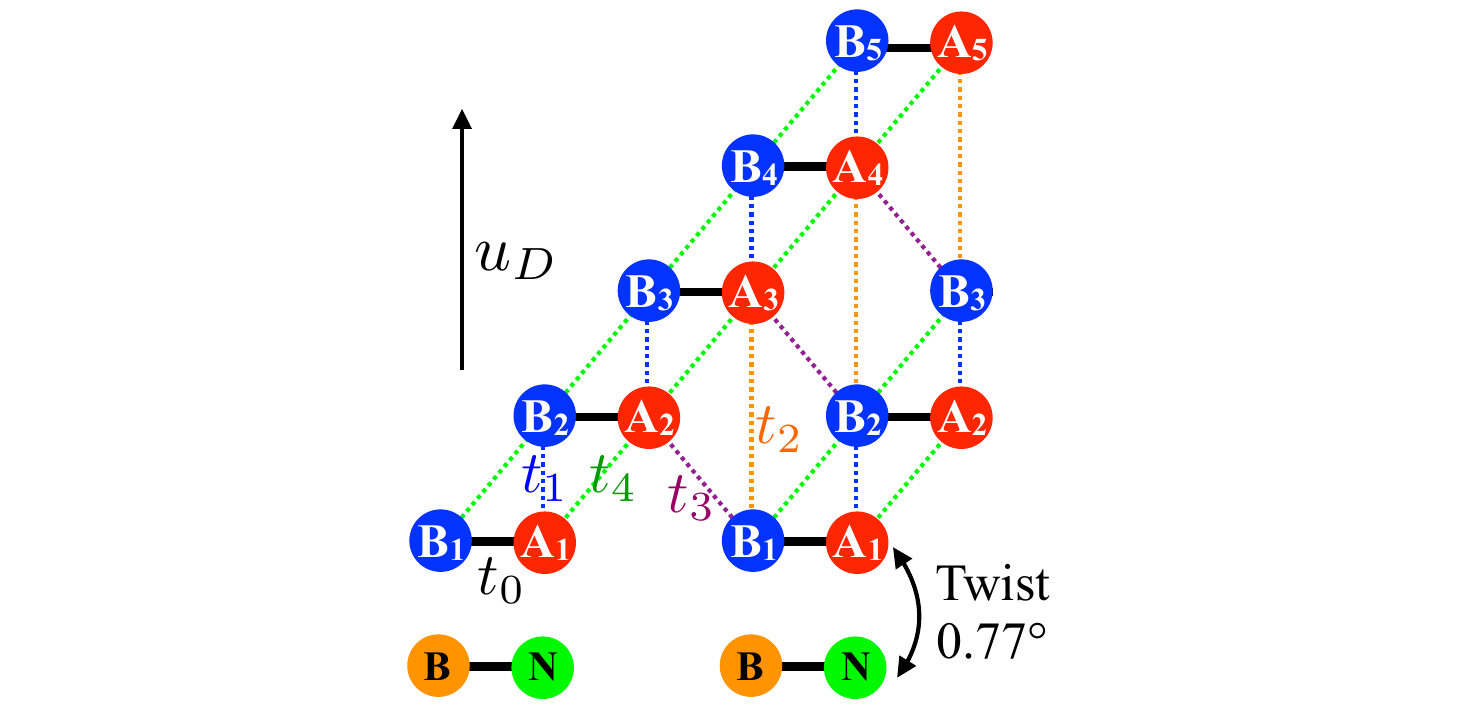}
 \caption{
 Pentalayer rhombohedral graphene on hBN layer. 
 $A_l$ and $B_l$ are the sublattices of the $l$th layer. The 
 solid and dotted lines represent intralayer $t_0$ 
 and interlayer $(t_1,t_2,t_3,t_4)$
 hopping terms. The twist angle between graphene and hBN is
 $0.77^\circ$. A perpendicular displacement field induces the interlayer 
 potential difference $u_D$.
 }
 \label{fig:graphene}
\end{figure}

The tight-binding Hamiltonian of pentalayer rhombohedral graphene
is given by~\cite{Zhang10,Park23-2,Dong23,JDong23}
\begin{align}
 \tilde{h}_\text{R5G}(\vec{k})=
 \left(
 \begin{array}{ccccc}
  D_{1} & V & W & 0 & 0 \\
  V^\dagger & D_{2} 
   & V & W & 0 \\
  W^\dagger & V^\dagger 
   & D_{3} & V & W\\
  0 & W^\dagger & V^\dagger 
   & D_{4} & V \\
  0 & 0 & W^\dagger & V^\dagger 
   & D_{5}
 \end{array}
 \right)
\end{align}
where
\begin{align}
 D_{l}
 &=\left(
 \begin{array}{cc}
  0 & t_0g(\vec{k}) \\
  t_0g^*(\vec{k}) & 0
 \end{array}\right)
 +u_D
 \left(l-3\right)\\
 V
 &=\left(
 \begin{array}{cc}
  t_4g^*(\vec{k}) & t_1 \\
  t_3g(\vec{k}) & t_4g^*(\vec{k})
 \end{array}
 \right)\\
 W
 &=\left(
 \begin{array}{cc}
  0 & 0 \\
  t_2 & 0
 \end{array}
 \right)\\
 g(\vec{k})
 &=\sum_{i=1}^3e^{-i\vec{k}\cdot\tau_i}.
\end{align}
Here, $t_0$ and $(t_1,t_2,t_3,t_4)$ are the intralayer and interlayer 
hopping terms [see Fig.~\ref{fig:graphene}], $u_D$ is the 
interlayer potential difference induced by a perpendicular displacement field,
and $\tau_i$'s are the vectors connecting the
nearest neighbor sublattices. The effective continuum Hamiltonian
$h_\text{R5G}(\vec{k})$ in the main text is given by replacing $g(\vec{k})$
with $a_\tx{G}\sqrt{3}/2(\xi k_x-ik_y)$, where $a_\tx{G}$ is the lattice 
constant of graphene and $\xi=+1$ $(-1)$ for the valley $K$ ($K'$).
Each parameter in Fig.~\ref{fig:graphene} is assigned as
$(t_0,t_1,t_2,t_3,t_4,u_D)=(-3100,380,-10.5,290,141,50)$meV, expected to
match experimental conditions for the $\nu=1$ QAH 
effect~\cite{Park23-2,Dong23,JDong23,ZLu24}.

\section{Moir\'e potential}
\label{sec:hBN}
We describe an effective continuum model for a \mo potential,
following an approach in Ref.~\onlinecite{Moon14}. 
Now, we consider the bottom layer of pentalayer rhombohedral graphene and 
hBN. The lattice constants of graphene and hBN are fixed as
$a_\tx{G}=0.246$nm and $a_\tx{hBN}=0.2504$nm~\cite{Liu03}, inducing the lattice
mismatch as $\epsilon=a_\tx{hBN}/a_\tx{G}-1\approx1.8\%$. The 
twist angle is set to $\theta=0.77^\circ$ to align with the experiment in 
Ref.~\onlinecite{ZLu24}. 

Let us first derive the primitive vectors of the \mo superlattice and the 
corresponding reciprocal lattice vectors. We denote the primitive lattice 
vectors of graphene by $\bm{a}_i$ with $i=1,2$. For hBN, we have
\begin{align}
 \tilde{\bm{a}}_i=MR\bm{a}_i,
\end{align}
where $M=(1+\epsilon)\bm{1}$, and $R$ is a rotation matrix by $\theta$. This
relation implies that a lattice of hBN at $\bm{r}$ has its counterpart of
graphene at $R^{-1}M^{-1}\bm{r}$. Their displacement is
\begin{align}
 \bm{\delta}(\bm{r})=(1-R^{-1}M^{-1})\bm{r}.
\end{align}
The primitive vectors of the \mo superlattice $\bm{L}_i^\tx{M}$ is defined so
that $\bm{\delta}(\bm{L}_i^\tx{M})=\bm{a}_i$:
\begin{align}
 \bm{L}_i^\tx{M}=(1-R^{-1}M^{-1})^{-1}\bm{a}_i.
\end{align}
In our settings, we have $|\bm{L}_i^\tx{M}|\approx45.4a_\tx{G}\approx11.2$nm.
The \mo reciprocal lattice vectors, satisfying 
$\bm{L}_i^\tx{M}\cdot\bm{G}_i^\tx{M}=2\pi\delta_{ij}$, are given by
\begin{align}
 \bm{G}_i^\tx{M}
 &=(1-M^{-1}R)\bm{a}_i^*\non
 &=\bm{a}_i^*-\tilde{\bm{a}}_i^*,
\end{align}
where $\bm{a}_i^*$ and $\tilde{\bm{a}}_i^*$ are the reciprocal lattice vectors 
of graphene and hBN, satisfying 
$\bm{a}_i\cdot\bm{a}^*_j=\tilde{\bm{a}}_i\cdot\tilde{\bm{a}}^*_j
=2\pi\delta_{ij}$. 
Here, the relation $\tilde{\bm{a}}_i^*=M^{-1}R\bm{a}_i^*$ is obtained by
$
\bm{a}_i\cdot\bm{a}^*_j
=\bm{a}_i^T\bm{a}^*_j
=(R^{-1}M^{-1}\tilde{\bm{a}}_i)^T\bm{a}^*_j
=\tilde{\bm{a}}_i^TM^{-1}R\bm{a}^*_j$.

The bilayer system composed of graphene and hBN is effectively described by
a tight-binding lattice Hamiltonian. Eliminating the hBN bases based on the
second order perturbation within an effective continuum framework, the effect 
of the hBN (for a given valley and spin)
is represented by a local potential $v(\bm{r})$ within the graphene 
subspace~\cite{Moon14}:
\begin{align}
 v(\bm{r})
 =&V_0
 \left(
 \begin{array}{cc}
  1 & 0 \\
   0 & 1
 \end{array}
 \right)\non
 &+\Bigg\{V_1e^{i\xi\psi}\bigg[
 \left(
 \begin{array}{cc}
  1 & \omega^{-\xi}\\
  1 & \omega^{-\xi}
 \end{array}
 \right)
 e^{i\xi\bm{G}_1^\tx{M}\cdot\bm{r}}
 +
 \left(
 \begin{array}{cc}
  1 & \omega^{\xi}\\
  \omega^{\xi} & \omega^{-\xi}
 \end{array}
 \right)
 e^{i\xi\bm{G}_2^\tx{M}\cdot\bm{r}}\non
 &+
 \left(
 \begin{array}{cc}
  1 & 1\\
  \omega^{-\xi} & \omega^{-\xi}
 \end{array}
 \right)
 e^{-i\xi\left(\bm{G}_1^\tx{M}+\bm{G}_2^\tx{M}\right)\cdot\bm{r}}
 \bigg]
 +\tx{H.c.}\Bigg\},
\end{align}
where $\omega=\exp\{2\pi i/3\}$ and $\xi=+1$ $(-1)$ for the valley 
$K$ ($K'$). Here, we use $V_0=28.9$meV, $V_1=21.0$meV and $\psi=-0.29$rad 
following Ref.~\onlinecite{Moon14}. As mentioned in the main text,
its second quantized form is 
\begin{align}
 V_\text{hBN}
 =\sum_{\vec{k}}\sum_{m_1m_2}\tilde{\bm{c}}^\dagger
 (\vec{k}+m_1\bm{G}_1+m_2\bm{G}_2)v(m_1,m_2)\tilde{\bm{c}}(\vec{k})
\end{align}
where $\tilde{\bm{c}}^\dagger=(c_{A_1}^\dagger,c_{B_1}^\dagger)$ and
$v(m_1,m_2)$ is the Fourier coefficient of $v(\vec{r})$. In the 
numerical 
calculations, the summation $\sum_{\vec{k}}\sum_{m_1m_2}$ is confined to the 
first and second \mo BZs. 

\section{Hartree-Fock calculation}
\label{sec:HF}
Let us describe the self-consistent HF calculation. As mentioned in the main 
text, the interaction before the mean-field 
approximation has the form of
\begin{align}
 &H_\text{int}
 =\frac{1}{2S}\sum_{\vec{k}\vec{k}'\vec{q}}\sum_{ZZ'}V_C(\vec{q})
 c^\dagger_Z(\vec{k}+\vec{q})c^\dagger_{Z'}(\vec{k}'-\vec{q})
 c_{Z'}(\vec{k}')c_Z(\vec{k}),
\end{align}
where $S$ is the area of the system, and $Z$ and $Z'$ represent the spin,
valley, sublattice and layer indices. 
Here, we set the positions of both $K$ and $K'$ to be the center in the 
folded \mo BZ for simplicity. 
The form of $V_C(\vec{q})$ is 
written in the main text. We then construct the Hartree and Fock
Hamiltonians as
\begin{align}
 H_\tx{H}
 &=\frac{1}{S}\sum_{\vec{k}\vec{k}'\vec{q}}\sum_{ZZ'}V_C(\vec{q})
 \left\langle c^\dagger_{Z'}(\vec{k}'-\vec{q})
 c_{Z'}(\vec{k}')\right\rangle\times\non
 &\qquad\qquad\qquad\qquad\qquad
 c^\dagger_Z(\vec{k}+\vec{q})c_Z(\vec{k}),
 \label{eq:HH}\\
 H_\tx{F}
 &=-\frac{1}{S}\sum_{\vec{k}\vec{k}'\vec{q}}\sum_{ZZ'}V_C(\vec{q})
 \left\langle c^\dagger_{Z'}(\vec{k}'-\vec{q})
 c_Z(\vec{k})\right\rangle\times\non
 &\qquad\qquad\qquad\qquad\qquad
 c^\dagger_Z(\vec{k}+\vec{q})c_{Z'}(\vec{k}'),
 \label{eq:HF}
\end{align}
where $\langle\cdot\rangle$ represents the expectation value for the ground 
state. 
Now, we pick up terms that couple momenta modulo the \mo reciprocal lattice 
vectors. 
In other words,
using a set of reciprocal lattice vectors of the \mo superlattice, denoted
$\bm{G}$,
we add constraints $\vec{k}+\vec{q}=\vec{k}+\bm{G}$ and 
$\vec{k}+\vec{q}=\vec{k}'+\bm{G}$ in Eqs.~\eqref{eq:HH} and \eqref{eq:HF},
respectively:
\begin{align}
 H_\tx{H}
 &\rightarrow\frac{1}{S}\sum_{\vec{k}\vec{k}'\bm{G}}\sum_{ZZ'}V_C(\bm{G})
 \left\langle c^\dagger_{Z'}(\vec{k}'-\bm{G})
 c_{Z'}(\vec{k}')\right\rangle\times\non
 &\qquad\qquad\qquad\qquad\qquad
 c^\dagger_Z(\vec{k}+\bm{G})c_Z(\vec{k}),
 \label{eq:tildeHH}\\
 H_\tx{F}
 &\rightarrow-\frac{1}{S}\sum_{\vec{k}\vec{k}'\bm{G}}\sum_{ZZ'}
 V_C(\vec{k}'-\vec{k}+\bm{G})
 \left\langle c^\dagger_{Z'}(\vec{k}-\bm{G})
 c_Z(\vec{k})\right\rangle\times\non
 &\qquad\qquad\qquad\qquad\qquad
 c^\dagger_Z(\vec{k}'+\bm{G})c_{Z'}(\vec{k}'),
 \label{eq:tildeHF}
\end{align}
In the numerical calculation, the summation $\sum_{\vec{k}\vec{k}'\bm{G}}$
is confined to the first and second \mo BZs. 
When seeking the ground state of the HF Hamiltonian for a given system 
parameters, we perform the self-consistent calculations with 10$-$20 
randomized initial states.
Specifically, we plug random numbers into 
$\langle c^\dagger_{Z'}(\vec{k}')c_Z(\vec{k})\rangle$ in the HF Hamiltonian
and solve them self consistently.
\begin{figure}[t!!]
\includegraphics[width=\columnwidth]{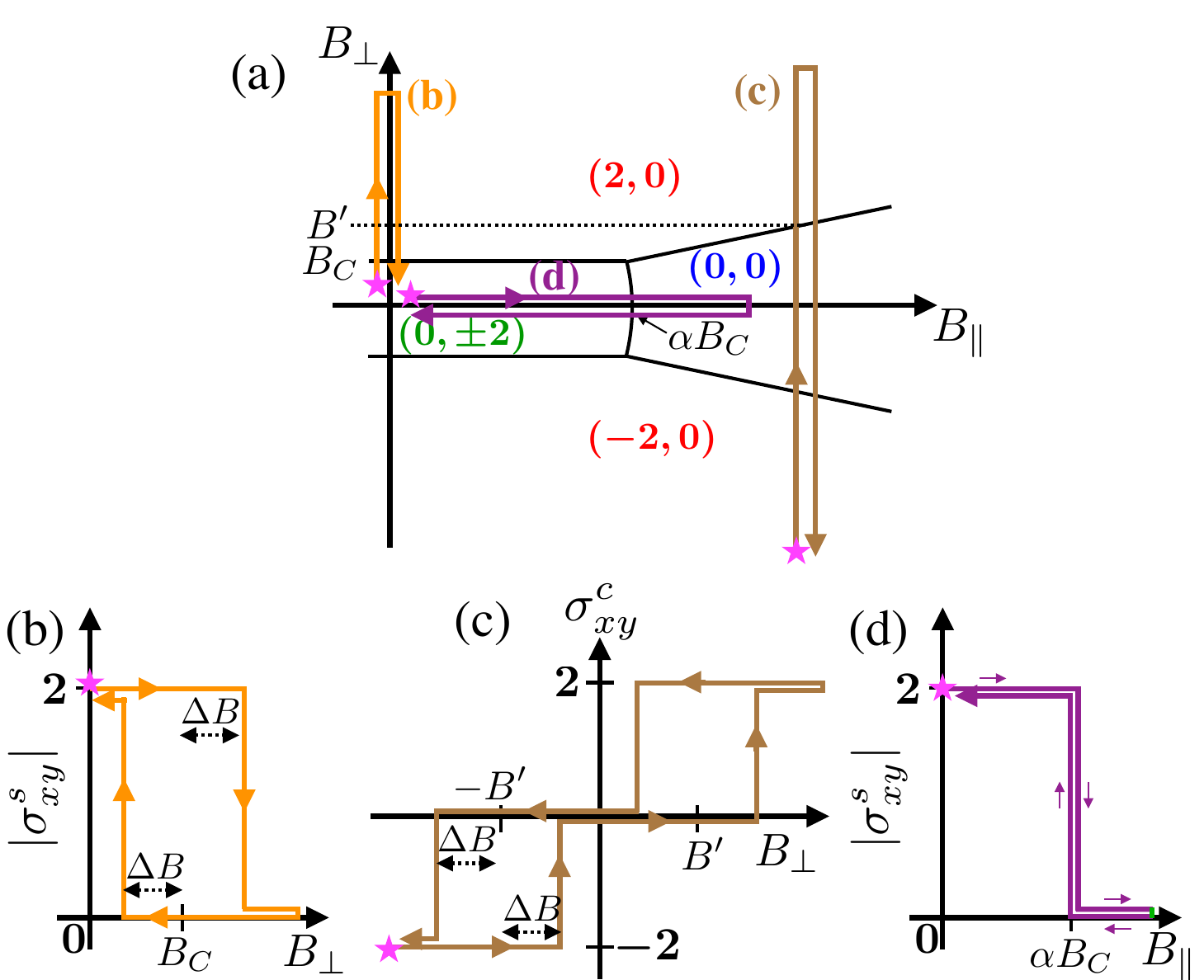}
 \caption{
 (a) The same as Fig.~\ref{fig:LE}. The three paths under consideration are
 illustrated, labeled by subsequent panels. The charge and spin Chern numbers 
 $(C_c,C_s)$ for each region are highlighted in red for QAH, green for QSH, and
 blue for QVH states. The magenta stars indicate the initial points of each
 path. (b)(c)(d) Charge and spin Hall conductances $\sigma_{xy}^c$ and 
 $|\sigma_{xy}^s|$. The coercive field in (b)(c) is represented by $\Delta B$.
 }
 \label{fig:hysteresis}
\end{figure}

\section{Hysteresis}
\label{sec:hysteresis}
We discuss hysteresis scans of the charge and spin Hall conductances 
$\sigma_{xy}^c$ and $\sigma_{xy}^s$. We consider three 
paths in the $(B_\perp,B_\para)$ space as illustrated in 
Fig.~\ref{fig:hysteresis}(a).
(In the QSH region of the figure, the spin Chern number $C_s$ is denoted as 
$\pm2$ to reflects the degeneracy of the states \#3 and \#4 in
Table~\ref{tab:nu2candidates}. Although either 
state should be favored due to weak spin-orbit interactions in real systems,
we simplify our argument by considering only $|\sigma_{xy}^s|$ below.
Note that the QSH state we consider carries nonzero valley Chern number as 
well~\cite{Islam16}.)

Figures~\ref{fig:hysteresis}(b)-(d) depicts anticipated behaviors of the Hall 
conductance. Figures~\ref{fig:hysteresis}(b) and (c) exhibit hysteresis since 
transitions from the QAH to QSH/QVH states involve valley flips, leading
to a first-order transition. The coercive field $\Delta B$ in
both figures is expected to be comparable to the experimentally observed one 
for the $\nu=1$ QAH effect~\cite{ZLu24} because the valley of only one band
 needs to be flipped. Conversely, Fig.~\ref{fig:hysteresis}(d) shows no
hysteresis behavior since transition from the QSH to QVH states involves only
spin flips. The SU(2) spin-rotational symmetry prevents hysteresis behavior
in such cases.

\begin{figure}[t]
\includegraphics[width=\columnwidth]{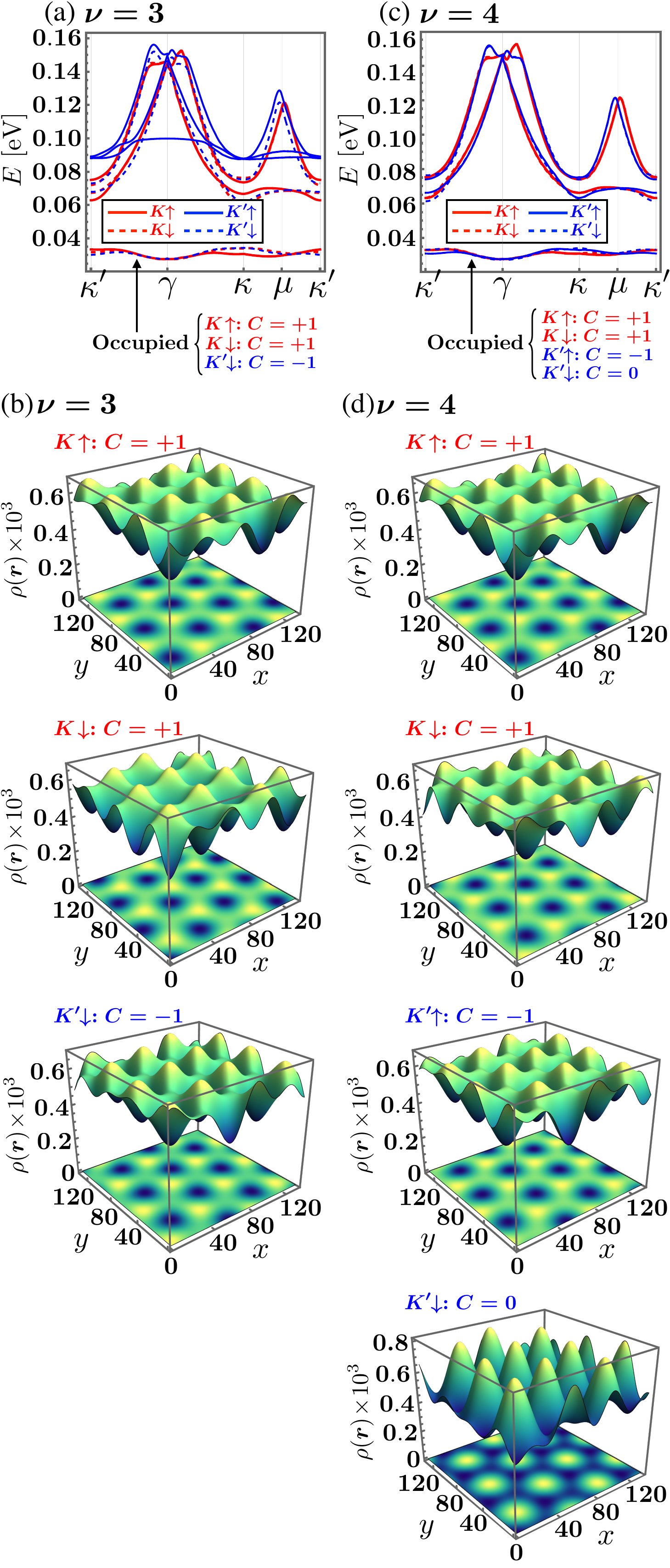}
 \caption{
 (a) HF band structure at $\nu=3$. The three lowest bands carry $C=+1$ or $-1$.
 (b) Charge densities $\rho(\vec{r})$ of the three lowest bands in (a).
 (c) HF band structure at $\nu=4$. One of the four lowest bands carries $C=0$.
 (d) Charge densities $\rho(\vec{r})$ of the four lowest bands in (c). 
 }
 \label{fig:WCL34}
\end{figure}
\section{Numerical results at $\nu=3$ and $4$}
\label{sec:WCL34}
Let us discuss the HF band structures and charge densities at $\nu=3$ and 4.
Figure~\ref{fig:WCL34}(a) presents the band structure of the lowest energy
state at $\nu=3$. The occupied
three lowest bands are separated from other conduction bands. Each band
carries the Chern number $C=+1$ or $-1$ depending on the valley. Their charge
densities $\rho(\vec{r})$ are shown in Fig.~\ref{fig:WCL34}(b). They provide
similar (but shifted) extended structure. 

Figure~\ref{fig:WCL34}(c) presents the band structure of the lowest energy
state at $\nu=4$. The occupied 
four lowest bands are separated from other conduction bands. Three bands carry
$C=+1$ or $-1$ while the other does $C=0$. Their charge densities are shown in
Fig.~\ref{fig:WCL34}(d).

\bibliography{biblio_fqhe}

\begin{thebibliography}{97}
\bibitem [{deg()}]{deg}%
  \BibitemOpen
  \href@noop {} {}\bibinfo {note} {While the QSH and QAH states in
  Fig.~\ref{fig:band} have slightly different energies, we numerically find
  that the two states become degenerate when the moir\'e potential is switched
  off. The symmetry that connects these states remains an open
  question.}\BibitemShut {Stop}%
\bibitem [{val()}]{valleyZeeman}%
  \BibitemOpen
  \href@noop {} {}\bibinfo {note} {Our HF Hamiltonian commutes with the
  ``valley Zeeman term", $H_{V}(\vec{k})=-b_{V}(\tau_z\otimes\sigma_0)$, where
  $\tau_z=\pm$ labels the valleys $K$ and $K'$, and $\sigma_0$ is the identity
  matrix acting in the spin space. This addition merely shifts the energy at
  each sector. Setting $b_V$ to be sufficiently small, we obtain
  Fig.~\ref{fig:band}(b). The additional energy shift [$=\pm
  b_z\times(\text{particle number})$] is excluded in the figure. In
  Fig.~\ref{fig:alpha}, we also use similar terms in order to obtain the target
  states, namely \#1, \#3, and \#5.}\BibitemShut {Stop}%
\end{thebibliography}

\end{document}